\documentclass[twocolumn,showpacs,pre,floatfix,superscriptaddress]{revtex4}

\usepackage{subfigure}
\usepackage[utf8]{inputenc}
\usepackage{comment}
\usepackage{color} 
\usepackage{tabularx} 
\usepackage{epsfig}
\usepackage{amsmath} 
\usepackage{amssymb} 
\usepackage{graphicx}
\usepackage{wasysym}
\usepackage{oldgerm}
\usepackage{psfrag}

\newcommand{\ti}{\tau^{|q|}_{\rm{int}}} 
\newcommand{\tiq}{\tau^{q}_{\rm{int}}}
\newcommand{\te}{\tau^{|q|}_{\rm{exp}}} 
\newcommand{\tr}{\tau_{\rm{RT}}}
\newcommand{\pjq}{P_J(q)}
\newcommand{\pqo}{I_J(q_0)}
\newcommand{\Ob}{\cal A}

\begin{document}
\bibliographystyle{revtex}

\title{Correlations between the dynamics of parallel tempering and the
free-energy landscape in spin glasses}

\author{Burcu Yucesoy}
\affiliation{Department of Physics, University of Massachusetts, Amherst,
Massachusetts 01003, USA}

\author{Jonathan Machta}
\affiliation{Department of Physics, University of Massachusetts, Amherst,
Massachusetts 01003, USA}

\author{Helmut G.~Katzgraber}
\affiliation{Department of Physics and Astronomy, Texas A\&M University, 
College Station, Texas 77843-4242, USA}
\affiliation{Theoretische Physik, ETH Zurich CH-8093 Zurich, Switzerland}

\date{\today}

\begin{abstract}

We present the results of a large-scale numerical study of the
equilibrium three-dimensional Edwards-Anderson Ising spin glass with
Gaussian disorder.  Using parallel tempering (replica exchange) Monte
Carlo we measure various static, as well as dynamical quantities, such
as the autocorrelation times and round-trip times for the parallel
tempering Monte Carlo method. The correlation between static and dynamic
observables for 5000 disorder realizations and up to 1000 spins down to
temperatures at 20\% of the critical temperature is examined.  Our
results show that autocorrelation times are directly correlated with the
roughness of the free-energy landscape.

\end{abstract}

\maketitle

\section{Introduction}

Systems in statistical physics with rough free-energy landscapes, such
as spin glasses, are notoriously difficult to equilibrate using Markov
chain Monte Carlo methods. The basic reason for this difficulty is that
extremely long times are required to surmount free-energy barriers and,
thus, to efficiently and fully explore the minima in the free-energy
landscape. The parallel tempering algorithm (also known as ``replica
exchange Monte Carlo'') \cite{geyer:91,marinari:92,hukushima:96}, as
well as other multicanonical techniques \cite{janke:98,berg:99}, have
partially overcome this problem. However, there is little understanding
of the relevant time scales for these methods. The purpose of this work
is to shed light on the equilibration time of the parallel tempering
algorithm and to understand how it is related to equilibrium properties
such as the roughness of the free-energy landscape. To this end, we
apply parallel tempering to a large ensemble of disorder realizations of
the three-dimensional Edwards-Anderson (EA) Ising spin glass
\cite{edwards:75,binder:86}. Our main results are that equilibration
times vary significantly from one disorder realization to another and
that this variation is correlated with the structure of the free-energy
landscape.  The fact that equilibration times are broadly distributed
has been previously noted in a different context (see, for example,
Refs.~\cite{alder:04} and \cite{alvarez:10a}) but the extent of
variation was not fully appreciated because different measures of
equilibration time were used in past studies. The correlation of dynamic
time scales for parallel tempering and equilibrium properties has not
been previously recognized, again because of the use of insensitive
measures of dynamic time scales. In this study we compute
autocorrelation functions of observables to measure equilibration times
whereas previous studies have primarily relied on the round-trip time
and related quantities. Although our study is focused on parallel
tempering and the Edwards-Anderson Ising spin-glass model, we believe
that our qualitative results are likely to be relevant both to other
disordered systems with rough free-energy landscapes, as well as other
multicanonical simulation techniques.

Parallel tempering was independently introduced by Geyer
\cite{geyer:91} and Marinari and Parisi \cite{marinari:92}, as well as
Hukushima and Nemoto \cite{hukushima:96}.  Furthermore, an earlier
algorithm with many of the same essential features was introduced by
Swendsen and Wang \cite{swendsen:86}. In parallel tempering many copies
of the system at different temperatures are simulated in parallel. Each
copy (henceforth referred to as replica) is simulated using a standard,
single-temperature Monte Carlo scheme such as the Metropolis or heat
bath algorithms \cite{krauth:06}. The set of temperatures is chosen so
it includes both high temperatures (where rapid equilibration is
possible using the single-temperature Monte Carlo method) and low
temperatures of interest (where equilibration is not feasible using the
standard algorithm). In addition to the single-temperature Monte Carlo
moves, there are also {\em replica exchange} moves in which replicas at
neighboring temperatures swap their temperatures.  Replica exchange
moves permit replicas to diffuse in temperature space
\cite{katzgraber:06a}.  The diffusion of replicas in temperature space
permits an enormous speed-up relative to single-temperature Monte Carlo.
Instead of directly surmounting the high free-energy barriers present at
low temperatures,  barriers are indirectly surmounted via the following
process: A replica in one free-energy well at the lowest temperature
diffuses to the highest temperature where it easily moves to another
well and then diffuses back to the lowest temperature. This process is
called a ``round trip'' and it allows parallel tempering to reduce
exponential time scales for surmounting free-energy barriers to 
power-law time scales \cite{machta:09}.

In the context of simple models with rough free-energy landscapes it has
recently been shown \cite{machta:09} that the equilibration time of
parallel tempering depends strongly on the structure of the free-energy
landscape.  If there is a single dominant minimum in the landscape and
no first-order transition within the range of temperatures, then the
equilibration time is short. On the other hand, the presence of several
nearly degenerate minima or the existence of a first-order transition
leads to much longer equilibration times. These results in simple model
landscapes prompted us to look for similar phenomena in the
Edwards-Anderson model where different disorder realizations may have
quite different free-energy landscapes. In this study, rather than
measuring the free-energy landscape directly, we used the spin overlap
(order parameter) distribution as a proxy. The overlap distribution is
closely related to the free-energy landscape but is substantially easier
to measure in simulations.  Based on the results from simple model
free-energy landscapes, we hypothesize that when the overlap
distribution has a complex structure, implying many free-energy
minima the time scale for parallel tempering will tend to be longer than
when the overlap distribution is simple. Our results confirm this
hypothesis.

This study is part of a larger project whose goal is to understand the
nature of the low-temperature phase of finite-dimensional spin glasses.  The
low-temperature phase of the Edwards-Anderson model is poorly understood and
hotly debated. There are several competing theories for the
low-temperature behavior of the model
\cite{fisher:86,fisher:88,parisi:79,mezard:87,marinari:98,newman:96,newman:98,krzakala:00,palassini:00,katzgraber:01}
and even though many large-scale investigations have been conducted
\cite{palassini:00,katzgraber:01,katzgraber:02,katzgraber:06,alvarez:10},
a conclusive theory correctly explaining all phenomena has not been
agreed on. Our recent simulations \cite{yucesoy:12} point toward a
two-pure-state picture such as the droplet picture
\cite{mcmillan:84b,fisher:86,fisher:87,fisher:88,bray:86} or the chaotic
pairs picture \cite{newman:92,newman:96,newman:98}.  One motivation for
this work was to verify that the ensemble of disorder realizations used
in Ref.~\cite{yucesoy:12} was indeed equilibrated.  Therefore, we have
measured both equilibration times and overlap distributions for all
disorder realizations and correlated these quantities.  We also
demonstrate that the conservative use of the equilibration criterion
introduced in Ref.~\cite{katzgraber:01} is sufficient to ensure that
nearly all disorder samples are equilibrated.

Our results suggest both a method for improving parallel tempering and a
warning when using it in spin-glass simulations. On the one hand, we
find that many disorder realizations have quite short equilibration
times.  Thus it might be useful to implement an adaptive scheme where
some disorder realization are simulated for shorter times than other
realizations. State-of-the-art parallel tempering simulations of
equilibrium spin glasses require huge amounts of CPU time because
of the difficulty of reaching equilibrium and the need for a large
ensemble of disorder realizations. Thus, such an adaptive scheme has the
potential for large savings in CPU time.  On the other hand, the fact
that the overlap distribution (a quantity directly related to the
controversy surrounding the low-temperature phase of spin glasses) is
significantly correlated with the dynamics of the algorithm serves as a
warning that one must be extremely careful to ensure that essentially
all samples are well equilibrated in order to avoid systematic errors in
measuring disordered-averaged equilibrium properties.

The paper is organized as follows: In Sec.~\ref{sec:model} we introduce
the model, the parallel tempering algorithm, as well as the measured
observables. In Sec.~\ref{sec:results} we show our results, followed by
conclusions.

\section{Model and Methods}
\label{sec:model}

\subsection{The Edwards-Anderson Ising spin glass}
\label{subsec:model}

We study the three-dimensional EA Ising spin-glass
model, defined by the energy function
\begin{equation}
 H=-\sum_{\langle i,j\rangle} J_{ij} s_i s_j .
\end{equation} 
The sum is over the nearest neighbors on a simple cubic lattice with
periodic boundary conditions and side length $L$.  The Ising spins
$s_{i}$ take values $\pm1$ and the interactions $J_{ij}$ are quenched
random couplings chosen from a Gaussian distribution with zero mean and
unit variance.

\subsection{Parallel tempering Monte Carlo}
\label{subsec:pt}

Parallel tempering, also known as replica exchange Monte Carlo,
is a powerful tool for simulating systems with rough free-energy
landscapes \cite{earl:05,katzgraber:06a,machta:09} with
applications across many disciplines. To date, it is the most efficient
method for simulating spin glasses and other disordered systems in
more than two dimensions at low temperatures where dynamics are slow.

The algorithm works as follows: Several copies (hereafter also referred
to as {\em replicas}) of the system with the same disorder are simulated
in parallel.  Each replica is simulated at a different temperature using
a standard Markov chain Monte Carlo (MCMC) technique such as the
Metropolis or heat-bath algorithms
\cite{newman:99,krauth:06,katzgraber:09e}. In this work we use the heat
bath algorithm because of its better performance at low temperature. The
set of temperatures is chosen to span both the low temperatures of
interest, where equilibration is not feasible using single-temperature
MCMC methods, and high temperatures where equilibration is fast. In
addition to the MCMC sweeps of each temperature, there are also {\em
replica exchange} moves between replicas. A proposed replica exchange
move involves swapping the temperatures of two replicas at neighboring
temperatures. A replica exchange move is accepted with probability $p$,
where
\begin{equation}
\label{eq:repexprob}
 p = \min [1 , e^{(\beta-\beta')(E-E')}] .
\end{equation}
Here $\beta = 1/T$ is the inverse temperature of one replica and
$\beta'$ the inverse temperature of the neighboring replica, and $E$ and
$E'$ are the corresponding energies of the two replicas. Equation
(\ref{eq:repexprob}) ensures that the entire Markov chain, including
both single-temperature MCMC sweeps and replica exchange moves satisfies
detailed balance and converges to equilibrium.  Replica exchange allows
replicas to diffuse from the lowest temperature to the highest
temperature and back again.  These {\em round trips} enhance mixing and
greatly reduce equilibration times for rough energy landscapes.

The parallel tempering algorithm has free parameters that include the
set of temperatures and the ratio of single-temperature sweeps to
replica-exchange sweeps. Optimizing parallel tempering therefore
requires an appropriate choice of these parameters. The choice of the
number of replicas $N_T$ and their temperatures involves the following
trade-off: If there are many closely-spaced temperatures, the energy
distribution between adjacent temperatures overlaps strongly and the
probability that a proposed replica exchange is accepted, see
Eq.~(\ref{eq:repexprob}), is large. This suggests that the more replicas
one uses, the better, apart from the extra computational work involved in
carrying out single-temperature sweeps of each replica. However, the
motion of replicas in temperature space is diffusive such that the time
scale for a round-trip scales approximately as the square of the number
of temperatures. At least for simple model systems, parallel tempering
is optimized if the number of temperatures scales as the square root of
the number of spins \cite{machta:09}. The average time to complete one
round trip is often used to characterize the performance of parallel
tempering. Choosing a set of temperatures that minimizes the round trip
time is one of the ways proposed to optimize parallel tempering
\cite{katzgraber:06a,trebst:06,bittner:08,hasenbusch:10}.

\subsection{Simulation parameters}
\label{subsec:simparams}

\begin{figure}[h]
 \centering
\includegraphics[scale=1.]{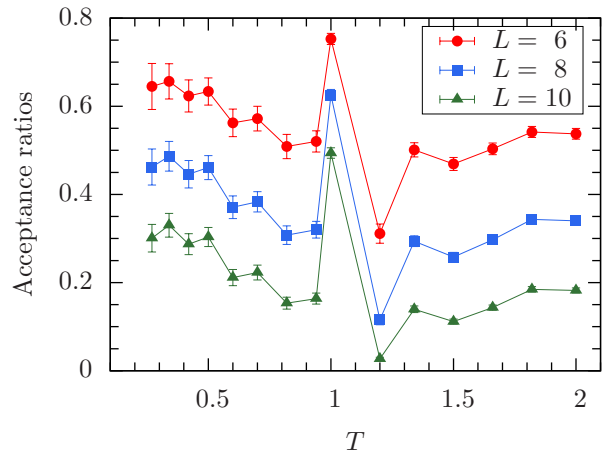}
\caption{(Color online) 
Average acceptance fraction as a function of temperature $T$ for replica
exchange moves. Note that $T_c \approx 0.96$.
}
\label{fig:AR}
\end{figure}

For the parallel tempering algorithm we use $N_T = 16$ temperatures
between $T = 0.2$ to $2.0$ \cite{comment:temp}.  This set of
temperatures was chosen heuristically in Refs.~\cite{katzgraber:01} and
\cite{katzgraber:02} to perform well for $L=6$ and $8$.  The
corresponding average acceptance fractions for replica exchange moves
are plotted in Fig.~\ref{fig:AR}. It is likely that the simulations would
be more efficient if more replicas were used near and above the critical
region ($T_c \approx 0.96$ \cite{katzgraber:06}) where the acceptance
fractions are small. However, we believe that as long as the swapping
probability is nonzero, our results will not change qualitatively. A
bottleneck merely slows down the diffusion process but does not prevent
it.  For $L=10$, the small acceptance fractions suggest that a larger
number of temperatures might be more efficient.  However, our goal is to
understand correlations between static and dynamic quantities in
parallel tempering for a given system size and so we have not sought to
optimize the algorithm for each system size.  We believe that the
qualitative results obtained using our parallel tempering parameters
would also hold for other reasonable values of the parameters.

By placing the hottest system temperature at $T \approx 2 T_c$, we
ensure that equilibration happens quickly, and we use one exchange sweep
for each heat-bath sweep ($N = L^3$ attempted spin updates). To
calculate the spin overlap [Eq.~(\ref{overlap})] we use two copies of
the system at each temperature; therefore, for each sample, we simulate
two sets of replicas independently from each other.

As stated before, we employ one replica exchange sweep after one 
heat-bath sweep. A heat-bath sweep corresponds to sequentially attempting to
update each spin at each replica once using the heat-bath algorithm. An
exchange sweep corresponds to randomly choosing $N_T - 1$ pairs of
adjacent replicas and proposing exchanges for each. Both types of sweeps
together make up one parallel tempering sweep. When no confusion arises,
we will call this unit of time simply a {\em sweep}.

For each system size we simulate approximately $5000$ disorder
realizations or {\em samples}. The simulations are equilibrated for at
least $2^{24}$ sweeps for $L = 6$ and $2^{27}$ sweeps for $L=8$ and
$10$, and then measurements are done for the same number of sweeps; see
Table \ref{tab:paramsea} for details. A fraction of samples required
longer runs to meet the equilibration criteria discussed in
Sec.~\ref{subsec:timescale}.  For samples with very short equilibration
times, we performed runs to obtain fine-grained autocorrelation
information; see Sec.~\ref{subsec:timescale}. In total, the data
collection took approximately 140 CPU years on 12-core AMD Opteron 6174
CPUs.

\begin{table}
\caption{
For each system size $L$ we equilibrate and then measure for at least
$2^b$ Monte Carlo sweeps. $T_{\rm min}$ ($T_{\rm max}$) is the lowest
(highest) temperature used, $N_T$ is the number of temperatures, and
$N_{\rm sa}$ is the number of disorder realizations. For some $L = 10$
samples longer runs had to be performed to ensure equilibration. 
\label{tab:paramsea}}
\begin{tabular*}{\columnwidth}{@{\extracolsep{\fill}} c r r r r r}
\hline
\hline
$L$ & $b$  & $T_{\rm min}$ & $T_{\rm max}$ & $N_{T}$ & $N_{\rm sa}$ \\ 
\hline
$6$  & $24$ & $0.2$         & $2.0$         & $16$    & $4961$ \\ \\
$8$  & $27$ & $0.2$         & $2.0$         & $16$    & $5126$ \\
$8$  & $28$ & $0.2$         & $2.0$         & $16$    &    $4$ \\ \\
$10$ & $27$ & $0.2$         & $2.0$         & $16$    & $4360$ \\
$10$ & $28$ & $0.2$         & $2.0$         & $16$    &  $353$ \\
$10$ & $29$ & $0.2$         & $2.0$         & $16$    &  $241$ \\
$10$ & $30$ & $0.2$         & $2.0$         & $16$    &   $73$ \\
\hline 
\hline
\end{tabular*}
\end{table}

\subsection{Observables}
\label{subsec:obs}

\subsubsection{Overlap distributions}
\label{subsec:distro}

\begin{figure}
\begin{center}
\subfigure{\includegraphics[scale=0.95]{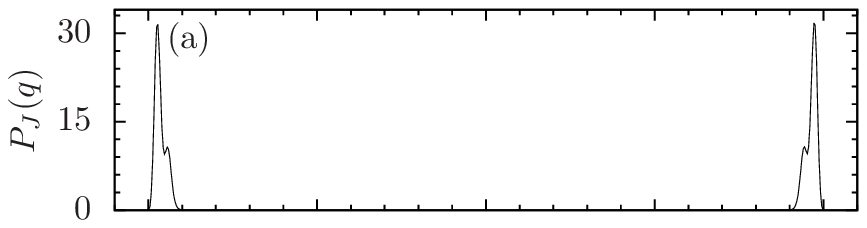}} 

\vspace*{-.38cm}    

\subfigure{\includegraphics[scale=0.94]{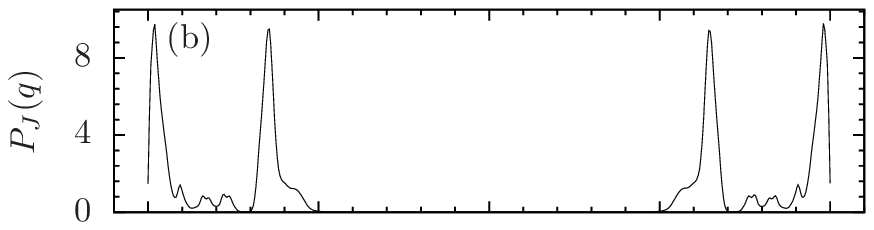}}
  
\vspace*{-.38cm}    

\subfigure{\includegraphics[scale=0.94]{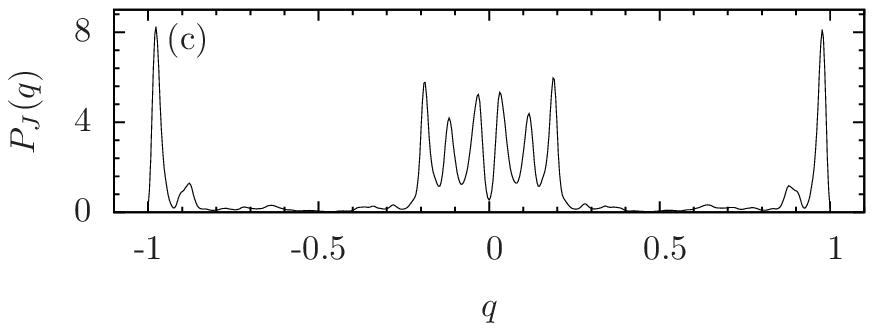}}
\end{center}
\caption{
Examples of overlap distributions $\pjq$ for three different disorder
realizations $J$ for $L=8$. While the distribution in panel (a) has only
two peaks, the distribution in panel (c) implies a very complex
free-energy landscape. All panels have the same horizontal scale.
}\label{OverlapDist}
\end{figure}

The focus of the paper is to relate static equilibrium properties of
individual spin-glass samples to the dynamics of the parallel tempering
algorithm acting on that sample. The primary quantity that we measure to
study both equilibrium properties and the dynamics of parallel tempering
is the spin overlap $q$,
\begin{equation}
q=\frac{1}{N}\sum^{N}_{i=0}s_i^{(1)}s_i^{(2)},
\label{overlap}
\end{equation}
where $N=L^3$ is the number of spins and the superscripts $(1)$ and
$(2)$ indicate two independent copies of the system with the same
disorder. The thermal and disorder average of the overlap is an order
parameter for the system and the probability distribution of the
thermally averaged overlap $\pjq$ for a given sample $J$ reveals aspects
of the free-energy landscape for that particular sample.  The behavior
of $\pjq$ for large systems and at low temperatures is one of the major
open questions in the theory of spin glasses. Overlap distributions vary
widely from sample to sample, as can be seen in the three examples shown
in Fig.~\ref{OverlapDist}. Although there is no direct mapping 
between the free-energy landscape and $\pjq$, it is clear that numerous
peaks in $\pjq$ imply numerous minima in the free-energy landscape. For
example, samples corresponding to Figs.~\ref{OverlapDist}(b) and
\ref{OverlapDist}(c) have a more complicated free-energy landscapes than
Fig.~\ref{OverlapDist}(a).  Our central hypothesis is that samples with
a more complicated free-energy landscape tend to have longer dynamic
time scales.

From $\pjq$ we define $\pqo$ for different $q_0$,
\begin{equation}
\label{eq:pqo}
\pqo = \int_{-q_0}^{q_0} \pjq ,
\end{equation} 
where $\pqo$ is the weight of $\pjq$ in an interval near the origin and it
serves as an approximate measure of the complexity of $\pjq$. For
example, $I_J(0.2) \approx 0$ for the samples shown in
Figs.~\ref{OverlapDist}(a) and \ref{OverlapDist}(b), while $I_J(0.2)$ is
large for the sample shown in Fig.~\ref{OverlapDist}(c). We compute
$\pqo$ for eight values of $q_0$ from $q_0=0.2$ to $0.9$.  It is worth
noting that $\pqo$---especially with $q_0=0.2$---has been extensively used
in studies of the low-temperature equilibrium properties of the EA model
\cite{katzgraber:01,alvarez:10}.

\subsubsection{Characteristic time scales}
\label{subsec:timescale}

We measure two time scales for parallel tempering from the
autocorrelation function of the overlap at the lowest temperatures. For
an arbitrary observable ${\Ob}$, the autocorrelation function
$\Gamma_{{\Ob}}(t)$ is defined via
\begin{equation}
\Gamma_{{\Ob}}(t) =
\frac{\langle {\Ob}(0){\Ob}(t)\rangle-\langle
{\Ob}\rangle^2}{\langle {\Ob}^2\rangle-
\langle {\Ob}\rangle^2}.
\end{equation}
Here, ${\Ob}(t)$ is the observable measured at time $t$ and $\langle
\cdots \rangle$ represents a thermal average \cite{comment:tauto}.
Integrated and exponential autocorrelation times are computed from the
autocorrelation function.  The integrated autocorrelation time
$\tau^{\Ob}_{\rm{int}}$ is the integral of the autocorrelation function
or, for discrete time, the sum of the autocorrelation function,
\begin{equation}
\tau^{\Ob}_{\rm{int}}=\frac{1}{2}
+\displaystyle\sum_{t=1}^{\infty}\Gamma_{{\Ob}}(t).
\label{tau}
\end{equation}
The integrated autocorrelation time is the time needed for two
subsequent measurements of ${\Ob}$ to decorrelate. It is a {\em lower
bound} on the time needed to reach equilibrium from an arbitrary initial
condition. To calculate $\tau^{\Ob}_{\rm{int}}$ care must be taken
when truncating the sum in Eq.~(\ref{tau}) to avoid large statistical
errors (explained below).

\begin{figure}
\begin{center}
\subfigure{\label{Auto-a}\includegraphics[scale=1.]{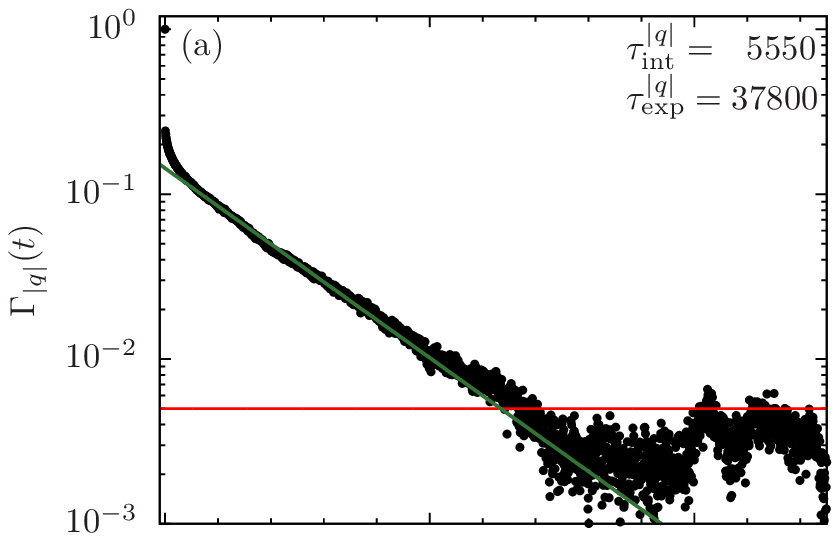}}  
    
\vspace*{-.38cm}    

\subfigure{\label{Auto-b}\includegraphics[scale=1.]{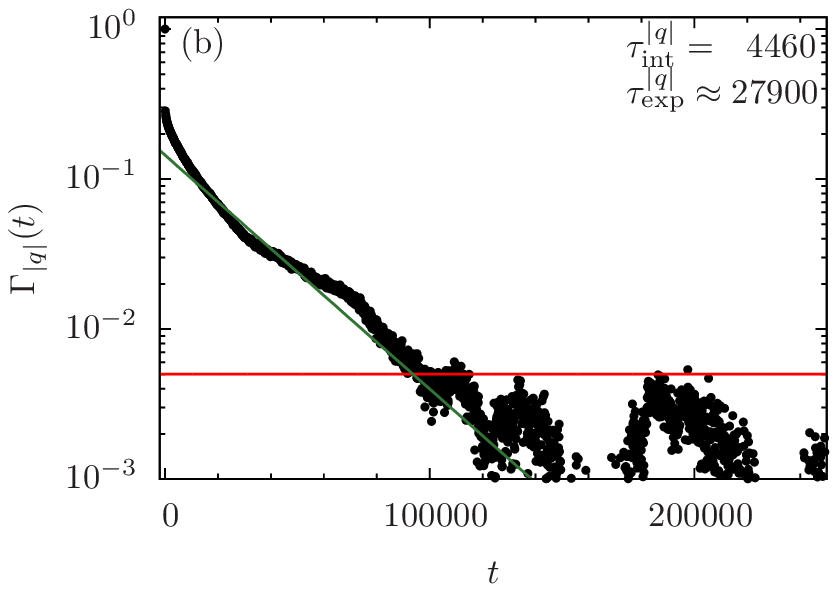}}
\end{center}
\caption{(Color online) 
Autocorrelation function for the absolute value of the order parameter
$\Gamma_{|q|}(t)$ as a function of simulation time for two different
disorder realizations and system size $L = 8$. The horizontal (red)
solid line at $\Gamma_{|q|} = 0.005$ represents the noise floor. The
diagonal solid (green) lines represent exponential fits to the data.
Both panels have the same horizontal scale.
}
\label{Auto}
\end{figure}

Figure \ref{Auto} shows two typical examples of the autocorrelation
function for the absolute value of the spin overlap $|q|$. For large
times, the autocorrelation function is eventually dominated by noise.
The noise floor is indicated in Figs.~\ref{Auto} by the horizontal (red)
solid line. Because the noise floor is determined primarily by the
number of data points used to compute the autocorrelation function, it
can be chosen to be the same for all simulations of the same length. Our
truncation procedure is to sum the autocorrelation function until it
first hits the noise floor. We obtain the noise floor for each run
length by visual inspection of $\Gamma_{|q|}(t)$ for many samples. We
note that this approach introduces some slight systematic bias in our
estimates of the integrated autocorrelation times. However, the main
emphasis of this paper is to understand the very large sample-to-sample
variations in time scales and, compared to these variations, the
errors introduced by our truncation protocol are small.

We have computed autocorrelation functions for both $|q|$ and $q$ and
from these we have obtained $\ti$ and $\tau^q_{\rm{int}}$, respectively.
To do this, we first compute a Fourier transform of the data and then
invert the function to obtain the autocorrelation function. This method
is considerably faster than a direct calculation.  For $L=6$, the
simulations are run for $2^{24}$ sweeps, recording data for the
computation of the autocorrelation functions every $10$ sweeps.  The
noise floor for these runs is set to $\Gamma = 0.005$. Some samples have
$\ti$ values of order 10 or less. For these, shorter data collection
runs of $2^{18}$ sweeps have been done starting from an equilibrated
spin configuration stored at the end of the longer run.  For these short
runs data are collected every sweep to accurately measure the
autocorrelation function up to time 10 and the full autocorrelation
function is patched together from the short and long runs. The noise
floor for these shorter runs is set to $\Gamma = 0.01$. If the noise
floor is reached before $t=10$, only the autocorrelation function
generated by the short run is used.  For both $L=8$ and $L=10$ the
simulations are run for at least $2^{27}$ sweeps, recording data every
$100$ sweeps with shorter runs also needed for samples with small $\ti$
values. The noise floor used is the same as for $L=6$.

The autocorrelation function is always calculated up to a maximum time
lag of 1\% of the total run length. Therefore, for some samples and, in
particular for $L=10$, the noise floor was not reached by this time and
longer runs were necessary to obtain good statistics. For these $L = 10$
samples we simulate up to $2^{30}$ sweeps. 32 samples (approximately
$0.6\%$) stayed above the noise floor even for the longest runs.  To
prevent a biasing of the results, these were not included in the
analysis.  However, it would be interesting to study these samples in
detail in the future.

There is another method for truncating the sum in Eq.~(\ref{tau})
introduced by Madras and Sokal \cite{madras:88} where the upper limit of
the sum is determined recursively. An initial upper limit is chosen and
$\tau^{\Ob}_{\rm{int}}$ is computed for that value, and the new upper
limit is determined as some factor ($6$ is suggested in
Ref.~\cite{madras:88}) times the current $\tau^{\Ob}_{\rm{int}}$. The
estimated value of $\tau^{\Ob}_{\rm{int}}$ is obtained when this
process has converged.  We experimented with this approach but did not
find that it converged in a consistent way across all samples with very
different autocorrelation functions. However, we note that our approach
should have systematic errors no larger than those resulting from the
method of Madras and Sokal with the upper limit set to $6\ti$ because
the noise floor is not reached until at least $10\ti$ except for the
very few samples with extremely small $\ti$ values.

The second time scale that we consider is the exponential
autocorrelation time $\tau^{\Ob}_{\rm{exp}}$, defined by the
asymptotic exponential decay of the autocorrelation function,
\begin{equation}
\Gamma_{{\Ob}}(t) \sim A e^{-t/\tau^{\Ob}_{\rm{exp}}}.
\end{equation}
Except for observables orthogonal to the slowest mode of the Markov
chain, $\tau^{\Ob}_{\rm{exp}}$ is the characteristic time of the
slowest mode and is thus the exponential time scale to reach equilibrium
from an arbitrary initial state \cite{sokal:89}.  

In principle, $\tau^{\Ob}_{\rm{exp}}$ is the most important time scale
for studying disordered systems because the main difficulty is reaching
thermal equilibrium (controlled by $\tau^{\Ob}_{\rm{exp}}$) rather
than collecting enough uncorrelated data at equilibrium (controlled by
$\tau^{\Ob}_{\rm{int}}$).  Unfortunately, calculating $\tau^{\cal
A}_{\rm{exp}}$ for a large number of samples is difficult because it
requires an automatized fitting process. For some samples, such as the
one shown in Fig.~\ref{Auto-a}, the autocorrelation function decays
nearly exponentially over most of the observable time range, allowing for
a precise fitting and extraction of $\ti$.  However, for many samples
the autocorrelation function is {\em not} exponential for the measured
times, as can be seen in Fig.~\ref{Auto-b} and, for these samples, an
automatic fitting procedure would not be reliable. For that reason, the
time scale that we correlate with the static properties of the EA spin
glass is the integrated autocorrelation time. We have studied $\te$ by
hand for a small subset of the data set and find that for those samples
where a reasonable fit is possible, $\te$ is in the range of $1$ to
$15$ times $\ti$.  Most, but not all, samples share the behavior seen in
the examples in Figs.~\ref{Auto} that there is an initial sharp decline
before the (approximately) exponential decay sets in. This sharp decline
explains why $\te$ is typically larger than $\ti$.

Finally, we also measured the round-trip time $\tr$ for each sample,
defined as the average time after equilibration is achieved, for a
replica to diffuse from the lowest to the highest temperature and then
back to the lowest temperature again \cite{katzgraber:06a}.

\section{Results}
\label{sec:results}

\begin{figure}
 \centering
\includegraphics[scale=1.]{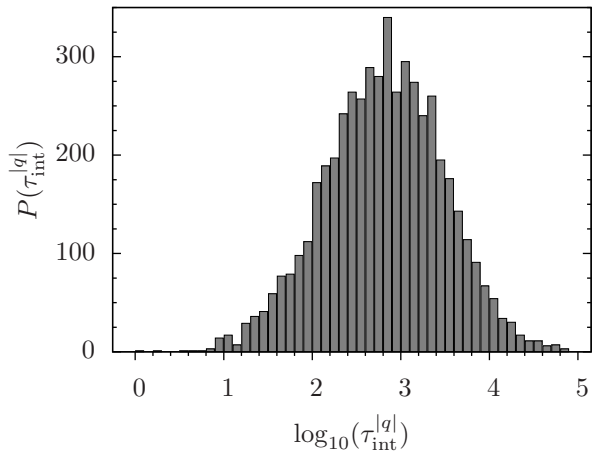}
\caption{
Logarithmic histogram of the integrated autocorrelation times $\ti$ for
for $L=8$. The data are normally distributed and show that a small
fraction of samples is extremely difficult to equilibrate.
}
\label{Histogram}
\end{figure}

\begin{figure}
\centering
\includegraphics[scale=1.]{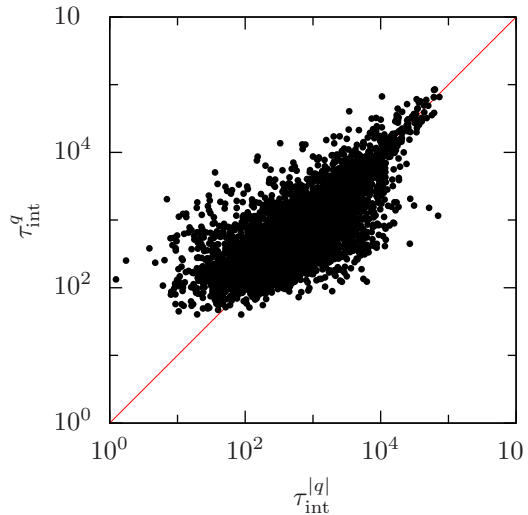}
\caption{(Color online)
Scatter plot of $\tiq$~vs.\ $\ti$ for $L=8$. The diagonal (red) solid
line corresponds to $\tiq=\ti$. Note that $\ti$ is often smaller than
$\tiq$ because the decorrelation of $q$ requires global spin flips while
the decorrelation of $|q|$ does not.
}
\label{QQ}
\end{figure}

Figure \ref{Histogram} shows the probability distribution
for $\ti$ for $L=8$.  In agreement with previous studies
\cite{alder:04,katzgraber:06,alvarez:10a}, the figure reveals that the
equilibration times for spin glasses are very broadly distributed.
The distributions for $L=6$ and $10$ are similar although with
different logarithmic means and standard deviations, as shown in
Tables \ref{tab:qmean} and \ref{tab:qstddev}, respectively.  Tables
\ref{tab:qmean} and \ref{tab:qstddev} also show the logarithmic
mean and standard deviation of $\tiq$.  The values for $\tiq$ are
similar to those for $\ti$, however, the standard deviations are
somewhat smaller. Figure \ref{QQ} shows a scatter plot with each point
representing one sample for $L = 8$. The $x$ and $y$ coordinates are
determined by  $\tiq$ and $\ti$, respectively.  The data illustrate
that $\tiq$ and $\ti$ are strongly correlated, especially for those
samples with the longest integrated autocorrelation times. The
observation that $\ti$ can take smaller values than $\tiq$ is
presumably a consequence of the fact that the decorrelation of $q$
requires global spin flips while the decorrelation of $|q|$ does not.

\begin{table}
\centering
\begin{subtable}
\centering
\caption{Mean values of the logarithms of $\tiq$, $\ti$ and $\tr$ for system
sizes $L = 6$ -- $10$.}
\label{tab:qmean}
\begin{tabular*}{\columnwidth}{@{\extracolsep{\fill}} l l l l}
\hline
\hline
$L$               & $6$         & $8$           & $10$        \\
\hline
$\log_{10}(\tiq)$ & $1.7164(58)$ & $2.8366(75)$ & $3.8711(84)$ \\
$\log_{10}(\ti)$  & $1.6035(90)$ & $2.8065(94)$ & $3.8881(97)$ \\
$\log_{10}(\tr)$  & $3.6217(28)$ & $4.4571(34)$ & $5.2421(42)$ \\ 
\hline
\hline
\end{tabular*}
\end{subtable}
\begin{subtable}
\centering
\caption{Standard deviations of the logarithms of $\tiq$, $\ti$ and $\tr$ for
system sizes $L = 6$ -- $10$.}
\label{tab:qstddev}
\begin{tabular*}{\columnwidth}{@{\extracolsep{\fill}} l l l l}
\hline
\hline
$L$               & $6$         & $8$           & $10$        \\
\hline
$\log_{10}(\tiq)$ & $0.408$     & $0.534$       & $0.592$     \\
$\log_{10}(\ti)$  & $0.631$     & $0.672$       & $0.685$     \\
$\log_{10}(\tr)$  & $0.194$     & $0.242$       & $0.293$     \\
\hline
\hline
\end{tabular*}
\end{subtable}

\end{table}

\begin{figure}
\begin{center}
\subfigure{\label{Scatter-a}\includegraphics[scale=0.9]{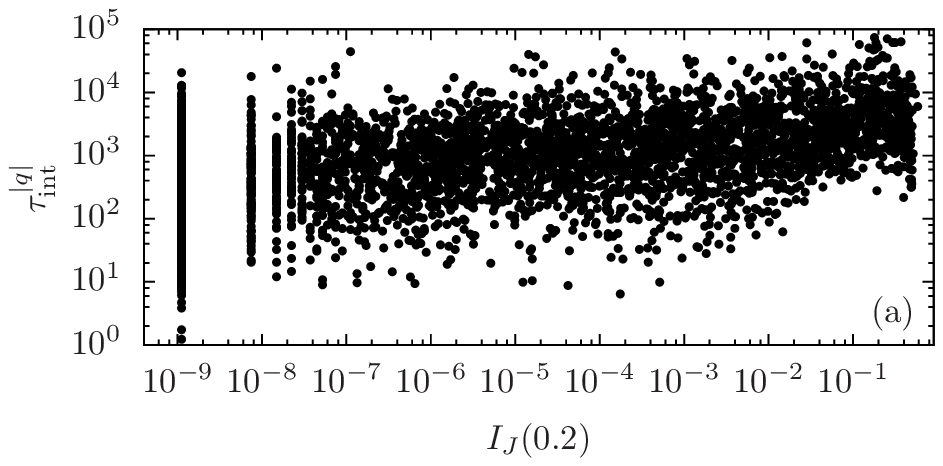}}
\subfigure{\label{Scatter-b}\includegraphics[scale=0.9]{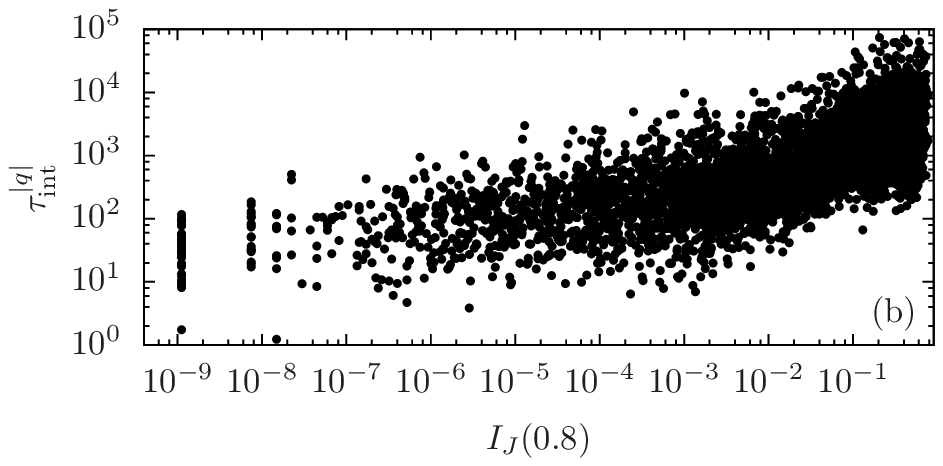}}
\end{center}
\caption{
Scatter plot for all $L=8$ samples of $\ti$ vs $\pqo$  for (a) $q_0=0.2$
and (b) $q_0=0.8$. The empty area underneath the scatter for large
$\pqo$ clearly shows that realizations with large values of $\pqo$ can
never have small values of $\ti$.
}
\label{Scatter}
\end{figure}

Figures \ref{Scatter}(a) and \ref{Scatter}(b) show scatter plots of
$\ti$ vs $\pqo$ for $L = 8$ and $q_0=0.2$ and $q_0=0.8$, respectively.
Each point represents one sample. These figures show that $\ti$ is
correlated with $\pqo$ and the correlation is stronger for larger
values of $q_0$. It is also striking that large values of $\pqo$
are never associated with very small integrated autocorrelation times
(similar results are obtained for the other system sizes studied). This
suggest that when the free-energy landscape is complex, time-scales
are large, as one would naively expect.

\begin{figure}[h]
\centering
\includegraphics[scale=1.]{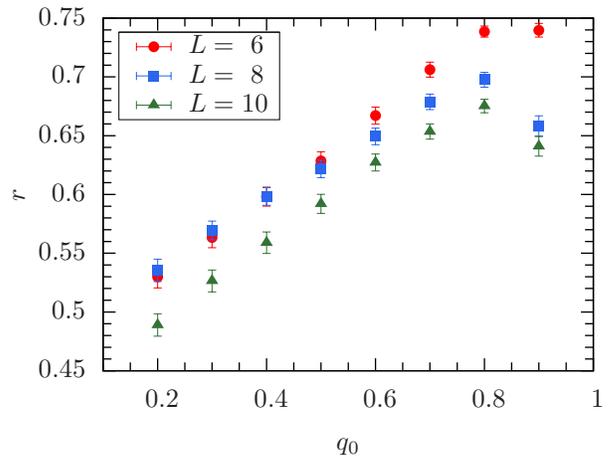}
\caption{(Color online)
Pearson correlation $r$ between $\log_{10}[\pqo]$ and $\log_{10}(\ti)$
vs $q_0$.  The data suggest that longer parallel tempering time scales
are correlated with more complicated overlap distributions.
}\label{corr2}
\end{figure}

To quantify the correlations between $\ti$ and $\pqo$, we use the
Pearson correlation coefficient $r$ \cite{press:95}, whose estimator for
a finite data set $\{(x_i,y_i)\}$ is
\begin{equation}
r = \frac{\displaystyle\sum_{i=0}^M(x_i-\bar{x})(y_i-\bar{y})}{\sqrt
{\displaystyle\sum_{i=0}^M(x_i-\bar{x})^2\displaystyle\sum_{i=0}^M
(y_i-\bar{y})^2}},
\end{equation}
where $M$ is the size of the data set and the overbar indicates the
average over the data set. If $r=1$ there is an exact linear
relationship between $x$ and $y$ while a small value of $r$ indicates
the absence of any linear dependence. Pearson $r$ values for
$\log_{10}(\ti)$ vs $\log_{10}[\pqo]$ for our data are shown in Table
\ref{tab:results} and Fig.~\ref{corr2} for the system sizes studied and
different values of $q_0$. These results clearly demonstrate that longer
parallel tempering time scales are correlated with more complicated
overlap distributions.  The strength of the correlation is slightly
weaker for larger systems.  The fact that $r$ increases with $q_0$
suggests that the presence of additional peaks in $\pjq$ increases $\ti$
independent of whether those peaks are near the origin.

\begin{table}
\caption{
Pearson correlation $r$ between the overlap weight $\log_{10}[\pqo]$ for
various $q_0$ and the logarithm of the integrated autocorrelation time
for the overlap $\log_{10}(\ti)$ for the three system sizes, $L$. The
bottom row shows the correlation coefficient between the logarithm of the
round trip time $\log_{10}(\tr)$ and $\log_{10}(\ti)$.
\label{tab:results}}
\begin{tabular*}{\columnwidth}{@{\extracolsep{\fill}} l l l l}
\hline
\hline
$q_0$            & $L = 6$      & $L = 8$      & $L = 10$    \\ 
\hline
$0.2$            & $0.5299(94)$ & $0.5355(95)$ & $0.4889(95)$\\ 
$0.3$            & $0.5634(86)$ & $0.5693(79)$ & $0.5264(93)$\\ 
$0.4$            & $0.5982(80)$ & $0.5983(75)$ & $0.5590(91)$\\ 
$0.5$            & $0.6286(77)$ & $0.6217(73)$ & $0.5920(80)$\\ 
$0.6$            & $0.6671(72)$ & $0.6494(70)$ & $0.6272(72)$\\ 
$0.7$            & $0.7062(63)$ & $0.6788(66)$ & $0.6535(65)$\\ 
$0.8$            & $0.7385(46)$ & $0.6975(63)$ & $0.6752(57)$\\ 
$0.9$            & $0.7397(58)$ & $0.6583(85)$ & $0.6410(83)$\\ 
\hline
$\log_{10}(\tr)$ & $0.014(19)$& $0.114(17)$~& $0.126(20)$~\\ 
\hline 
\hline
\end{tabular*} 
\end{table}  

\begin{figure}
\centering
\includegraphics[scale=1.]{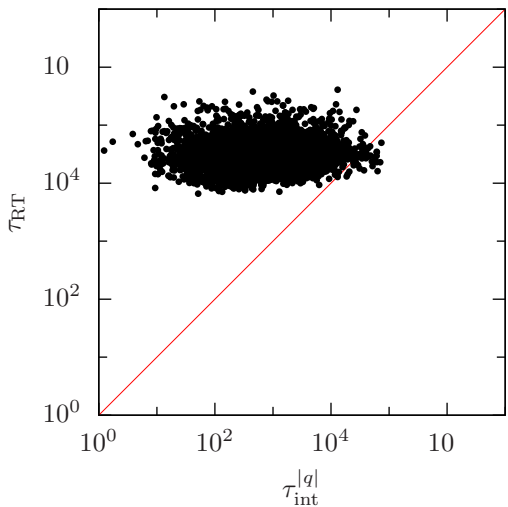}
\caption{(Color online)
Scatter plot of the average round trip time $\tr$ vs $\ti$ for $L=8$.
The diagonal (red) line is $\tr=\ti$. The data show almost no
correlation between these quantities.}
\label{RT}
\end{figure}

On the other hand, we find that the round trip time is not correlated
with $\ti$ or, by extension, $\pqo$. Figure \ref{RT} is a scatter plot
of $\tr$ vs $\ti$ for $L=8$ (compare with Fig.~\ref{QQ}).  Each point
represents one sample and the diagonal line is $\tr=\ti$. It is clear
from this figure that $\tr$ has a much smaller variance than $\ti$ and
that the correlation between the two is minimal. The last row of Table
\ref{tab:results} shows the Pearson $r$ value between $\log_{10}(\tr)$
and $\log_{10}(\ti)$ and quantifies the lack of correlation between
$\tr$ and $\ti$.  Note that the authors of Ref.~\cite{alvarez:10}
consider a time scale related to the motion of replicas in parallel
tempering and also find no correlation with $\pqo$.

If the highest temperature is chosen such that its mixing time is
one heat bath sweep, then $\tr \geq \ti$ because there will be no
memory of the spin state of the replica between successive visits
to the lowest temperature, as can be seen in Tables \ref{tab:qmean}
and \ref{tab:qstddev}, as well as Fig. \ref{RT}. Figure \ref{RT}
shows that $\tr$ is  much larger than $\ti$ for most samples.
One explanation for this is that (for most samples) the spin state of
the coldest replica can be decorrelated by diffusion from the lowest
temperature to an intermediate temperature and back; full round trips
are not needed for decorrelation.

Note that if we repeat the study for $\tiq$ instead of $\ti$, the
correlation values would show the same qualitative trend as for $\ti$,
but they would, on average, be slightly smaller.

Finally, we also test equilibration with the method introduced by
Katzgraber {\em et al.}~in Ref.~\cite{katzgraber:01} for short-range
spin glasses with Gaussian disorder between the spins. In this method,
the average energy per spin
\begin{equation}
u = - \frac{1}{N} \sum_{\langle i,j \rangle} [\, J_{ij} \langle s_i s_j
\rangle \, ]_{\rm av} ,
\end{equation}
where $[ \cdots ]_{\rm av}$ represents a disorder average and, as before,
$\langle \cdots \rangle$ denotes the Monte Carlo average for a given
set of bonds, can be related to the link overlap
\begin{equation}
q_l = \frac{1}{N_b} \sum_{\langle i, j \rangle}
s_i^{(1)} s_j^{(1)} 
s_i^{(2)} s_j^{(2)}
\label{eq:ql}
\end{equation}
via an integration by parts over the interactions $J_{ij}$ between the
spins.  In Eq.~\eqref{eq:ql} $N_b = dN$ with $d = 3$ the space dimension 
represents the number of bonds in the system.  One obtains:
\begin{equation}
[\langle q_l \rangle]_{\rm av} =
1 - \frac{T|u|}{d} ,
\label{eq:equil_cond}
\end{equation}
where $T$ is the temperature.

\begin{figure}
\centering
\includegraphics[scale=1.]{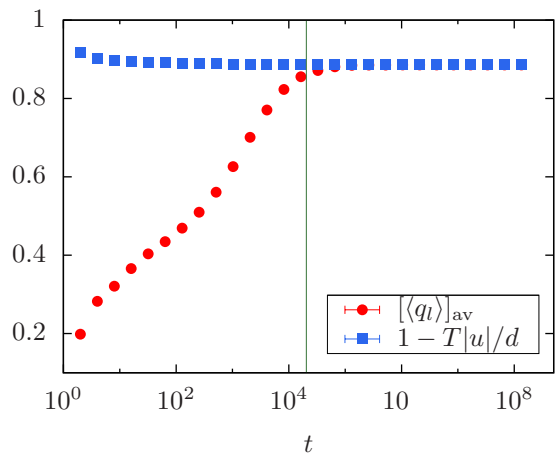}
\caption{(Color online)
Average link overlap $[\langle q_l \rangle]_{\rm av}$ (red circles)
and $1 - T|u|/d$ [see Eq.~\eqref{eq:equil_cond}, blue squares] as a
function of time $t$ for $L = 8$ and $T = 0.20$.  For $t \gtrsim 10^5$
sweeps both data sets agree, suggesting that the system is in thermal
equilibrium.  The vertical (green) line in the figure represents the
time for which 99\% of the samples have values of $\ti$ less than
this time. Error bars are smaller than the symbols.
}\label{fig:Eq}
\end{figure}

Figure \ref{fig:Eq} shows representative results for $L = 8$ and $T =
0.2$, the lowest temperature simulated. At approximately $10^5$ Monte
Carlo sweeps both data computed directly from the energy per spin
(blue squares) and data computed from the link overlap (red circles)
agree within error bars. Note that 99\% of the $\ti$ values for this
system size are smaller than the point marked with the solid vertical
(green) line in Fig.~\ref{fig:Eq}. Thus, for the simulated time of
over $10^8$ Monte Carlos sweeps, the data are well equilibrated.
This suggests that a conservative use of the equilibration criterion
developed in Ref.~\cite{katzgraber:01} will guarantee that almost
all of the samples are equilibrated.

\section{Conclusions}

We have found one factor that explains the broad distribution of time
scales in parallel tempering for spin glasses: The roughness of the
free-energy landscape of each individual sample directly affects its
equilibration time scales. In accordance with previous observations on
the parallel tempering method \cite{machta:09}, if there are many minima
in the free-energy landscape, a sample requires more time to
equilibrate.  The equilibrium distribution of the spin overlap serves as
proxy for the complexity of the free-energy landscape. Our results show
that individual samples need to be tested individually to ensure proper
equilibration, as previously suggested in Ref.~\cite{alvarez:10a}.
However, we show that the equilibration test developed in
Ref.~\cite{katzgraber:01} is a viable alternative for system sizes
currently accessible in simulations if applied conservatively.
Furthermore, the parallel tempering round-trip times seem to be
unaffected by the complexity of the free-energy landscape and should
therefore not be used as a measure of how well a system is equilibrated.

The clear correlations between algorithmic (nonequilibrium) time
scales and the equilibrium complexity of the free-energy landscape
opens the door for alternate studies of the nature of the spin-glass
state, a problem that remains controversial.

\begin{acknowledgments}

We appreciate useful discussions with J.~C.~Andresen and R.~S.~Andrist.
H.G.K.~would like to thank J.~C.~Andresen for making a superb
Tres Leches.  H.G.K.~acknowledges support from the Swiss National
Science Foundation (Grant No.~PP002-114713) and the National Science
Foundation (Grant No.~DMR-1151387). J.M.~and B.Y.~are supported in
part by the National Science Foundation (Grants No.~DMR-0907235 and
No.~DMR-1208046).  We thank the Texas Advanced Computing Center (TACC)
at The University of Texas at Austin for providing HPC resources
(Ranger Sun Constellation Linux Cluster), ETH Zurich for CPU time
on the Brutus cluster, and Texas A\&M University for access to their
eos and lonestar clusters.

\end{acknowledgments}

\bibliographystyle{apsrevtitle} 
\bibliography{refs,comments}

\begin{thebibliography}{42}
\bibitem[{com({\natexlab{b}})}]{comment:tauto}
\bibinfo{note}{To implement the used definition of the autocorrelation function
  in the simulations, the observable ${\cal A}$ is measured at regular
  intervals and correlated with its value at a time lag $t$ earlier. See, for
  example, Ref.~\cite{newman:99}.}
\end{thebibliography}

\end{document}